\documentclass[12pt]{amsart}
\usepackage{amssymb,latexsym,amsmath,enumitem,mathrsfs,geometry}
\usepackage{thmtools}
\usepackage{fullpage}
\usepackage{fancyhdr}
\usepackage{xcolor}
\usepackage{bbm}
\usepackage{stmaryrd}
\usepackage{mathrsfs}
\usepackage{todonotes}
\usepackage{hyperref}
\makeatletter
\renewcommand{\enddoc@text}{%
  \ifx\@empty\@translators\else
    \@settranslators
  \fi
  \ifx\@empty\addresses\else
    \clearpage
    \phantomsection
    \addcontentsline{toc}{section}{Authors' Affiliations}
    \@setaddresses
  \fi
}
\usepackage{eso-pic}
\usepackage{graphicx}
\usepackage{subcaption}
\usepackage{diagbox}
\usepackage{array}
\usepackage{tikz}
\usetikzlibrary{arrows}
\usetikzlibrary{positioning, arrows.meta, shapes.geometric}
\usepackage{float}
\usepackage{comment}
\usepackage{mathtools}
\usepackage{amsthm}
\usepackage{placeins}
\usepackage{acro}
\usepackage{setspace}
\usepackage{lineno}
\newtheorem{theorem}{Theorem}[section]
\newtheorem{definition}[theorem]{Definition}
\newtheorem{corollary}[theorem]{Corollary}
\newtheorem{proposition}[theorem]{Proposition}
\newtheorem{lemma}[theorem]{Lemma}
\newtheorem{remark}[theorem]{Remark}
\newtheorem{example}[theorem]{Example}
\DeclareAcronym{RV}{
short = RV,
long  = \emph{Random Variable},
tag   = abbrev}
\DeclareAcronym{KL}{
short = KL,
long  = \emph{Kullback--Leibler},
tag   = abbrev}
\DeclareAcronym{GM}{
short = GM,
long  = \emph{Gaussian Mixture},
tag   = abbrev}
\DeclareAcronym{PDF}{
short = PDF,
long  = \emph{Probability Density Function},
tag   = abbrev}
\DeclareAcronym{EM}{
short = EM,
long  = \emph{Expectation-Maximization},
tag   = abbrev}
\usepackage{cleveref}
\usepackage[
    backend=biber,
    style=numeric,
    sorting=none,
    doi=true,
    url=true,
    eprint=true,
    isbn=true
]{biblatex}
\begin{document}
\numberwithin{equation}{section}
\title{
    Closed-Form Gaussian Information Projection for Embedded Simplices and Simplicial Mixtures
    }
\author{Ali Darijani$^{\circledast\square\#\P\perp}$}
\address{
Ali Darijani$^{\circledast\square\#\P\perp}$\newline
\indent ORCID: \url{https://orcid.org/0009-0000-4659-093X}\newline
\indent Research Staff, Fraunhofer IOSB of Fraunhofer-Gesellschaft, Germany\newline
\indent Research Staff, KIT Department of Computer Science, Germany\newline
\indent Alumnus, RWTH Aachen University, Germany\newline
\indent Alumnus, Sharif University of Technology, Iran
}
\email{ali.darijani@iosb.fraunhofer.de}
\email{ali.darijani@kit.edu}
\email{ali.darijani@rwth-aachen.de}
\email{ali.darijani@alum.sharif.edu}
\author{Emadedin Soufian$^{\square\#\P\perp}$}
\address{
Emadedin Soufian$^{\square\#\P\perp}$\newline
\indent ORCID: \url{https://orcid.org/0009-0006-5777-4623}\newline
\indent Research Staff, Det Norske Veritas (DNV), Germany\newline
\indent Alumnus, RWTH Aachen University, Germany\newline
\indent Alumnus, Iran University of Science and Technology, Iran
}
\email{emadedin.soufian@dnv.com}
\email{emadedin.soufian@rwth-aachen.de}
\author{Jürgen Beyerer$^{\circledast\perp}$}
\address{
Jürgen Beyerer$^{\circledast\perp}$\newline
\indent ORCID: \url{https://orcid.org/0000-0003-3556-7181}\newline
\indent Director, Fraunhofer IOSB of Fraunhofer-Gesellschaft, Germany\newline
\indent Professor, KIT Department of Computer Science, Germany\newline
\indent Alumnus, Karlsruhe Institute of Technology, Germany
}
\email{juergen.beyerer@iosb.fraunhofer.de}
\email{juergen.beyerer@kit.edu}
\address{
Symbol Legend\newline
\indent${\circledast}$: Idea, ${\square}$: Math, ${\#}$: Programming, ${\P}$: Writing, ${\perp}$: Critique
}

\thanks{Funded by the Deutsche Forschungsgemeinschaft (DFG) - Project number 471687386}
\date{\today}
\begin{abstract}
We study Gaussian approximation of Dirichlet--Gaussian probability 
density functions on embedded simplices. We derive the unique information 
projection onto the family of nondegenerate ambient Gaussian probability density functions, 
obtaining closed-form expressions for the optimal Gaussian and the minimum divergence 
for arbitrary positive Dirichlet parameters. The minimum divergence depends only on 
the intrinsic simplex dimension and Dirichlet parameters, is invariant up to an 
affine isomorphism, independent of the ambient dimension, and independent of the 
normal covariance. Furthermore, by joint convexity of Kullback--Leibler divergence, 
we establish explicit intrinsic-dimension-dependent sharp upper bounds for finite 
Gaussian Mixture with shared component weights. For uniform probability density functions, 
these bounds reduce to dimension-dependent constants. The results provide closed-form Gaussian Mixture 
constructions and rigorous approximation guarantees for probability density functions 
associated with embedded simplicial complexes. A numerical example on a circular 
curve illustrates the theoretical results and their connections to geometric 
Gaussian Mixture representations.
\end{abstract}
\maketitle
\tableofcontents
\AddToShipoutPictureBG{
    \AtPageCenter{
        \begin{tikzpicture}[remember picture,overlay]
            \node[
                rotate=-45,
                text=gray!25
            ] at (current page.center) {
                \upshape\normalfont\resizebox{0.9\paperwidth}{!}{\bfseries Preprint}
            };
        \end{tikzpicture}
    }
}
\section{Introduction}
\label{sec:introduction}

Gaussian projection provides a canonical approximation of a \ac{PDF} by a 
Gaussian~\ac{PDF}~\cite{stuart_inverse,info_geo}, 
minimizing the \ac{KL} divergence. For \ac{PDF} associated with geometric 
simplicial structures~\cite{munkres_alg_top}, 
this problem admits additional structure through barycentric coordinates and the 
geometry of the embedding~\cite{darij_gm_curve,darij_gm_ieee,darij_simplex}.

We consider \acp{PDF} on embedded simplices generated by 
Dirichlet \acp{PDF} on the simplex tangent space and independent Gaussian \acp{PDF} on
the simplex normal space. This formulation handles uniform \ac{PDF} as well as non-uniform 
\acp{PDF} on simplices embedded in arbitrary finite ambient dimensions.

We derive the unique information projection~\cite{info_geo} onto the 
Gaussian family, which is characterized by moment matching~\cite{stuart_inverse,info_geo}, 
and obtain a closed-form expression for the information projection divergence. 
Remarkably, this quantity depends only on the intrinsic simplex dimension and 
the Dirichlet \ac{PDF} parameters, and is invariant under affine isomorphism (by construction), 
independently of the ambient dimension and normal \ac{PDF} a result similar to~\cite{darij_simplex}.

Furthermore, we extend the componentwise result to finite \acp{GM} of embedded simplicial 
\acp{PDF}. By joint convexity of the \ac{KL} divergence~\cite{info_geo}, we 
establish explicit intrinsic-dimension-dependent sharp upper bounds for the corresponding \ac{GM}. The uniform \ac{PDF} 
follows as a special case.
The main contributions are:
\begin{itemize}
    \item Gaussian information projection:
    Characterization of the unique optimal Gaussian for
    embedded Dirichlet--Gaussian \ac{PDF}.

    \item Closed-form divergence:
    Derivation of the information projection divergence and its
    invariance with respect to the affine isomorphism,
ambient dimension, and normal \ac{PDF}.

    \item \ac{GM} approximation:
    Establishment of explicit information projection sharp upper 
    bounds for finite \acp{GM} associated with embedded
    simplicial complex based \acp{PDF}.
\end{itemize}

The remainder of the paper develops the geometric and
measure theoretic probability framework, derives the 
componentwise information projection, and establishes 
the corresponding information projection \ac{GM} bounds.
\section{Previous Work}
\label{sec:previous_work}

This work generalizes our previous results on a \ac{GM} 
representations of triangulated domains
\cite{darij_simplex}, which considered
uniform~\ac{PDF} on full-dimensional simplices.
We extend that framework in two directions:
\begin{itemize}
    \item from uniform \acp{PDF} to arbitrary Dirichlet~\ac{PDF} 
    with positive parameters, and
    \item from full-dimensional simplices to simplices embedded 
    in arbitrary finite ambient dimensions with independent Gaussian \ac{PDF} on their normal spaces. 
\end{itemize}

These extensions give a general closed-form
characterization of the componentwise Gaussian
information projection and corresponding \ac{GM}
divergence bounds. The previous uniform model is
recovered as a special case.
\section{Construction}
\label{sec:construction}

We first introduce the geometric and measure theoretic probability notation required
to construct \ac{GM} representation of embedded simplicial
complexes. Throughout this paper, the ambient space is
$\mathbb R^D$, whereas the simplices may have intrinsic dimensions
$d\leq D$. Vectors are column vectors, $I_D$ denotes the identity
matrix on $\mathbb R^D$, and $A^\top$ denotes the transpose of a
matrix $A$. For a symmetric positive-definite matrix $\Sigma$,
$\mathcal N_D(x|m,\Sigma)$ denotes the Gaussian \ac{PDF} on
$\mathbb R^D$ with mean $m$ and covariance $\Sigma$.

\begin{definition}[Convex Hull~\cite{munkres_alg_top}]
Let $A\subseteq\mathbb R^D$. Its convex hull is defined by
\begin{equation}
    \operatorname{conv}(A)\coloneq\left\{\sum_{i=1}^{n}\lambda_i v_i:n\in\mathbb N_{>0}, v_i\in A, \lambda_i\geq0, \sum_{i=1}^{n}\lambda_i=1\right\}.
\end{equation}
\end{definition}

\begin{definition}[Standard Simplex~\cite{munkres_alg_top}]
The standard $d$-simplex is
\begin{equation}
    \Delta^d\coloneq\operatorname{conv}\{0,e_1,\ldots,e_d\}\subseteq\mathbb R^d,
\end{equation}
where $e_i$ denotes the $i$-th canonical basis vector.
Equivalently,
\begin{equation}
    \Delta^d=\left\{x\in\mathbb R^d:x_i\geq0,\quad\sum_{i=1}^{d}x_i\leq1\right\}.
\end{equation}
\end{definition}

\begin{definition}[Geometrically Independent Points~\cite{munkres_alg_top}]
$\{v_i\}_{i=0,\dots,d} \subset \mathbb{R}^D$ are geometrically independent iff \(\mathrm{rank}\left[v_1-v_0 \ldots v_d-v_0\right]=d\).
\end{definition}

\begin{definition}[Embedded Simplex~\cite{lee_topol}]
Let $v_0,\ldots,v_d\in\mathbb R^D$ be geometrically independent points; then
\begin{equation}
    \sigma^d=\operatorname{conv}\{v_0,\ldots,v_d\}\subseteq\mathbb R^D
\end{equation}
is called an embedded $d$-simplex.
\begin{remark}
    The intrinsic dimension of an embedded simplex is $d$, independently of the ambient
dimension $D$.
\end{remark}
\end{definition}

\begin{definition}[Faces of an Embedded Simplex]
Let
\begin{equation}
    \sigma^d=\operatorname{conv}\{v_0,\ldots,v_d\}
\end{equation}
be an embedded simplex.
A nonempty face of $\sigma^d$ is a simplex of the form
\begin{equation}
    \tau=\operatorname{conv}\{v_i:i\in I\},\qquad\varnothing\neq I\subseteq\{0,\ldots,d\}.
\end{equation}
Its dimension is
\begin{equation}
    \dim\tau=|I|-1,
\end{equation}
where \(|I|\) is the cardinality of the set \(I\).
In particular, its $0$-faces, $1$-faces, and
$(d-1)$-faces are called vertices, edges, and facets,
respectively.
\end{definition}

\begin{definition}[Simplicial Complex in $\mathbb R^D$]
\label{def:embedded-simplicial-complex}

A finite geometric simplicial complex
$\mathcal S$ embedded in $\mathbb R^D$ is a finite collection
of nondegenerate simplices satisfying:

\begin{enumerate}
    \item Every nonempty face of a simplex in $\mathcal S$
    also belongs to $\mathcal S$.

    \item For any two simplices $\sigma,\tau\in\mathcal S$,
    their intersection is either empty or a common face
    of both simplices.
\end{enumerate}
The geometric realization of $\mathcal S$ is
\begin{equation}
    |\mathcal S|\coloneq\bigcup_{\sigma\in\mathcal S}\sigma\subseteq\mathbb R^D.
\end{equation}
\end{definition}

\begin{definition}[Dimension and Purity]
\label{def:dimension-purity}

Let $\mathcal S$ be a finite simplicial complex.
Its intrinsic dimension is
\begin{equation}
    \dim\mathcal S\coloneq\max_{\sigma\in\mathcal S}\dim\sigma.
\end{equation}
A simplex $\sigma\in\mathcal S$ is called maximal if it is
not a proper face of any other simplex in $\mathcal S$.
The complex is called pure of dimension $d$ if every maximal
simplex has dimension $d$.
We denote the collection of its $d$-dimensional simplices by
\begin{equation}
    \mathcal S^{[d]}\coloneq\{\sigma\in\mathcal S:\dim\sigma=d\}.
\end{equation}
\end{definition}

\begin{definition}[Embedded Triangulation]
\label{def:embedded-triangulation}

Let $\Omega\subseteq\mathbb R^D$ be a set with
intrinsic dimension $d$ and finite positive $d$-dimensional
Hausdorff measure~\cite{evans_measure}.
A finite embedded triangulation of $\Omega$ is a finite,
pure, $d$-dimensional simplicial complex
$\mathcal S$ such that \(|\mathcal S|=\Omega.\)
We denote the intrinsic $d$-dimensional volume by
$\operatorname{Vol}_d$, which agrees with the
$d$-dimensional Hausdorff measure on the affine hull of
each simplex.
\end{definition}

\begin{proposition}[Intersections of Maximal Simplices]
\label{prop:maximal-simplex-intersection}

Let $\mathcal S$ be a finite, pure, $d$-dimensional simplicial complex embedded in $\mathbb R^D$.
For distinct $\sigma,\tau\in\mathcal S^{[d]}$,
\begin{equation}
    \sigma\cap\tau=\varnothing\quad\text{or}\quad\sigma\cap\tau\in\mathcal S,\quad\dim(\sigma\cap\tau)\leq d-1.
\end{equation}
Consequently,
\begin{equation}
    \operatorname{Vol}_d(\sigma\cap\tau)=0.
\end{equation}
\end{proposition}

\begin{proof}
By~\Cref{def:embedded-simplicial-complex},
the intersection of two simplices is either empty or a
common face. Since $\sigma$ and $\tau$ are distinct
$d$-simplices, their common face has dimension at most
$d-1$. Every lower-dimensional face has zero
$d$-dimensional Hausdorff measure.
\end{proof}

\begin{corollary}[Volume of an Embedded Triangulation]
\label{cor:volume-triangulation}

Let $\mathcal S$ be a finite, pure, $d$-dimensional simplicial complex embedded in $\mathbb R^D$.
Then
\begin{equation}
    \operatorname{Vol}_d(|\mathcal S|)=\sum_{\sigma\in\mathcal S^{[d]}}\operatorname{Vol}_d(\sigma).
\end{equation}
\end{corollary}

\begin{proof}
The geometric realization is the union of its maximal
$d$-simplices. By~\Cref{prop:maximal-simplex-intersection},
distinct maximal simplices intersect only in sets of
zero $d$-dimensional volume. Finite additivity of
Hausdorff measure therefore gives the result.
\end{proof}

\begin{proposition}[Affine Parametrization of an Embedded Simplex]
\label{prop:affine-embedded-simplex}

Let
\begin{equation}
    \sigma^d=\operatorname{conv}\{v_0,\ldots,v_d\}\subseteq\mathbb R^D
\end{equation}

be an embedded simplex.
Define its edge matrix
\begin{equation}
    E_d\coloneq
\begin{bmatrix}
v_1-v_0 & \cdots & v_d-v_0
\end{bmatrix}
\in\mathbb R^{D\times d}.
\end{equation}
Then the unique affine bijection from the standard simplex is
\begin{equation}
   F:\Delta^d \to \sigma^d,\quad x\mapsto E_dx+v_0.
\end{equation}
\end{proposition}

\begin{proof}
Since the vertices are geometrically independent,
$E_d$ has full column rank. Therefore, $F$ is injective.
Moreover,
\begin{equation}
    F(0)=v_0,\quad F(e_i)=v_i,\quad i=1,\ldots,d.
\end{equation}
By affinity, $F$ maps the convex hull of the vertices of
$\Delta^d$ onto the convex hull of the vertices of
$\sigma^d$. Hence $F$ is an affine bijection.
Uniqueness follows because an affine map is uniquely
determined by its values on affinely independent points.
\end{proof}

\begin{proposition}[Intrinsic Volume of an Embedded Simplex]
\label{prop:embedded-simplex-volume}

Let \(\sigma^d=\operatorname{conv}\{v_0,\ldots,v_d\}\)
be an embedded simplex in $\mathbb R^D$.
Its intrinsic volume is
\begin{equation}
    \operatorname{Vol}_d(\sigma^d)=\frac{1}{d!}\sqrt{\det(G_d)},
\end{equation}
where
\begin{equation}
    G_d\coloneq E_d^\top E_d\in\mathbb R^{d\times d}
\end{equation}
is the Gram matrix of its edge vectors.
In particular, $G_d$ is positive definite.
\end{proposition}

\begin{proof}
The affine parametrization in~\Cref{prop:affine-embedded-simplex}
has constant $d$-dimensional volume factor
\begin{equation}    
    J_d(F)=\sqrt{\det(E_d^\top E_d)}.
\end{equation}
Since
\begin{equation}
    \operatorname{Vol}_d(\Delta^d)=\frac{1}{d!},
\end{equation}
the change-of-variables formula for the affine map
gives the stated result.
\end{proof}

\begin{definition}[Barycentric Subdivision]
Let \(\sigma^d=\operatorname{conv}\{v_0,\ldots,v_d\}\)
be a simplex.
For each nonempty subset
$I\subseteq\{0,\ldots,d\}$, define its barycenter by
\begin{equation}
    b_I\coloneq\frac{1}{|I|}\sum_{i\in I}v_i.
\end{equation}
The barycentric subdivision of \(\sigma^d\) consists of
the simplices whose vertices are barycenters of nested
sequences of nonempty faces:
\begin{equation}
    b_{I_0},b_{I_1},\ldots,b_{I_r},\quad I_0\subsetneq I_1\subsetneq\cdots\subsetneq I_r.
\end{equation}
For a $d$-simplex, its maximal subdividing simplices
correspond to permutations $(i_0,\ldots,i_d)$ of
$\{0,\ldots,d\}$ and have vertices
\begin{equation}
    b_{\{i_0\}},b_{\{i_0,i_1\}},\ldots,b_{\{i_0,\ldots,i_d\}}.
\end{equation}
These simplices form a triangulation of the original
simplex.
\end{definition}

\begin{remark}    
Applying barycentric subdivision consistently to all
simplices of an embedded simplicial complex produces
a finer embedded simplicial complex without changing
its geometric realization. In particular, neighboring
simplices induce identical subdivisions on their common
faces. This construction allows local refinement of an embedded
mesh. The \ac{GM} representation can subsequently
be applied to the refined simplices using their respective
vertices, intrinsic dimensions, and intrinsic volumes.
\end{remark}

\begin{definition}[Simplex Tangent and Normal Spaces]
\label{def:tangent-normal}

Let
\begin{equation}
    \sigma^d=\operatorname{conv}\{v_0,\ldots,v_d\}\subseteq\mathbb R^D
\end{equation}
be an embedded simplex with $d<D$.
Its tangent space is
\begin{equation}
    T_\sigma\coloneq\operatorname{span}\{v_i-v_0:i=1,\ldots,d\}.
\end{equation}
Its normal space is the orthogonal complement
\begin{equation}    
    N_\sigma\coloneq T_\sigma^\perp=\{z\in\mathbb R^D:z^\top t=0\text{ for every }t\in T_\sigma\}.
\end{equation}
Consequently,
\begin{equation}
    \mathbb R^D=T_\sigma\oplus N_\sigma,\quad\dim T_\sigma=d,\quad\dim N_\sigma=D-d.
\end{equation}
The affine hull of the simplex is \(v_0+T_\sigma\).
\end{definition}

\begin{remark}[Covariance Operators on Tangent and Normal Spaces]
    \label{rem:covariance_operator_tanget_normal}

Throughout the paper, covariance operators on the tangent and normal
spaces are distinguished from their ambient matrix representations.

\noindent Let \(U_N\in\mathbb R^{D\times(D-d)}\) have 
orthonormal columns spanning \(N_\sigma\). 
The normal covariance operator is denoted by
\(\Sigma_N:N_\sigma\to N_\sigma\) and is positive definite on
\(N_\sigma\). Its matrix representation \(S_N\) in the
orthonormal basis \(U_N\) is the unique matrix satisfying
\begin{equation}
    \Sigma_N(U_Nz)=U_NS_Nz\quad\text{for every }z\in\mathbb R^{D-d},\quad S_N\in\mathbb S_{++}^{D-d}.
\end{equation}
The corresponding ambient representation, obtained by extending
\(\Sigma_N\) by zero on \(T_\sigma\), is denoted by
\begin{equation}
    \overline\Sigma_N\coloneq U_NS_NU_N^\top\in\mathbb S_+^D.
\end{equation}
Thus, \(\overline\Sigma_N\) is positive 
semidefinite on \(\mathbb R^D\), positive 
definite on \(N_\sigma\), and vanishes
on \(T_\sigma\). In particular, it is 
generally singular as an ambient \(D\times D\) 
matrix.
\end{remark}

\begin{definition}[Intrinsic Differential Entropy]
\label{def:intrinsic-entropy}

Let \(A=v_0+T_\sigma\) denote the affine hull of 
the embedded simplex \(\sigma^d\). We write 
\(\mathcal H^d|_A\) for the restriction of the 
\(d\)-dimensional Hausdorff measure to \(A\), 
with the standard normalization that agrees with 
ordinary Euclidean volume on \(d\)-dimensional 
affine spaces.

For an \ac{RV} \(Y\) supported on \(A\) and 
absolutely continuous with respect to 
\(\mathcal H^d|_A\), let \(f_Y\) denote its 
\ac{PDF} with respect to this measure. Its 
differential entropy is defined as

\begin{equation}
    h_{T_\sigma}(Y)\coloneq-\int_A f_Y(x)\log f_Y(x)\mathrm d\mathcal H^d(x),
\end{equation}
provided the integral is well-defined.
Equivalently, if \(U_T\in\mathbb R^{D\times d}\) 
has orthonormal columns spanning \(T_\sigma\), 
then \(h_{T_\sigma}(Y)\) equals the ordinary 
differential entropy of the coordinate 
representation \(U_T^\top(Y-v_0)\) in \(\mathbb R^d\).
Throughout the paper, differential entropies of 
\ac{RV} supported on \(d\)-dimensional 
affine subspaces are understood in this intrinsic 
sense. Differential entropies of absolutely 
continuous \ac{PDF} on \(\mathbb R^D\) are 
understood with respect to \(D\)-dimensional 
Lebesgue measure.

\end{definition}

\begin{definition}[Dirichlet \ac{PDF} on the Standard Simplex]
\label{def:dirichlet-simplex}

Let
\begin{equation}
    \Delta^d=\left\{x\in\mathbb R^d:x_i\geq0,\ \sum_{i=1}^dx_i\leq1\right\}.
\end{equation}
For $x\in\Delta^d$, define the barycentric coordinates
\begin{equation}
    x_0:=1-\sum_{i=1}^dx_i.
\end{equation}
Given
\begin{equation}
    \alpha_\sharp=(\alpha_0,\ldots,\alpha_d)^\top\in\mathbb R_{++}^{d+1},
\end{equation}
the Dirichlet \ac{PDF} is
\begin{equation}
    \mathcal D_{\Delta^d}(x|\alpha_\sharp)=\frac{1}{\mathcal{B}(\alpha_\sharp)}\prod_{i=0}^{d}x_i^{\alpha_i-1},
\end{equation}
where
\begin{equation}
    \mathcal{B}(\alpha_\sharp)=\frac{\prod_{i=0}^{d}\Gamma(\alpha_i)}{\Gamma(|\alpha_\sharp|_1)}.
\end{equation}
\end{definition}

\begin{corollary}[Uniform \ac{PDF} on the Standard Simplex]
\label{cor:uniform-standard-simplex}

For
\begin{equation}
    1_{d+1}\coloneq(1,\ldots,1)^\top\in\mathbb R^{d+1}_{++},
\end{equation}
the Dirichlet \ac{PDF} with parameter
$\alpha_\sharp=1_{d+1}$ is uniform on $\Delta^d$:
\begin{equation}
    \mathcal D_{\Delta^d}(x|1_{d+1})=
    \begin{cases}
        d!,     &x\in\Delta^d,\\
        0,      &x\notin\Delta^d.
    \end{cases}
\end{equation}
\end{corollary}

\begin{proposition}[Mean and Covariance of the Dirichlet \ac{PDF}]
\label{prop:dirichlet-moments}

Let \(X=(X_1,\ldots,X_d)^\top\) follow the Dirichlet \ac{PDF} 
with parameter \(\alpha_\sharp\in\mathbb R_{++}^{d+1}\).
Define
\begin{equation}
    \alpha_\natural\coloneq(\alpha_1,\ldots,\alpha_d)^\top,\quad a\coloneq|\alpha_\sharp|_1.
\end{equation}
Then
\begin{equation}
    \mathbb E(X)=\frac{\alpha_\natural}{a},\quad \mathbb{V}(X)=\frac{a\operatorname{diag}(\alpha_\natural)-\alpha_\natural\alpha_\natural^\top}{a^2(a+1)}.
\end{equation}
\end{proposition}

\begin{definition}[Dirichlet \ac{PDF} on an Embedded Simplex]
\label{def:dirichlet-embedded-simplex}

Let
\begin{equation}
    \sigma^d=\operatorname{conv}\{v_0,\ldots,v_d\}\subseteq\mathbb R^D, \quad 1\leq d<D,
\end{equation}
be an embedded simplex, and let
$F(x)=E_dx+v_0$ be its affine parametrization.
For \(\alpha_\sharp\in\mathbb R_{++}^{d+1}\),
the Dirichlet \ac{PDF} on \(\sigma^d\) is the
pushforward of the Dirichlet \ac{PDF} on $\Delta^d$
under $F$. Equivalently, if
\begin{equation}
    X\sim\mathcal D_{\Delta^d}(x|\alpha_\sharp),
\end{equation}
then
\begin{equation}
    Y\coloneq E_dX+v_0
\end{equation}
follows the Dirichlet \ac{PDF} on \(\sigma^d\).
We denote this \ac{PDF} by
$\mathcal D_{\sigma^d}(y|\alpha_\sharp)$.
\end{definition}
\begin{proposition}[Mean and Covariance on an Embedded Simplex]
\label{prop:embedded-dirichlet-moments}

Under~\Cref{def:dirichlet-embedded-simplex},
the mean and covariance of $Y$ are
\begin{equation}
    m_T=\mathbb E(Y)=v_0+E_d\frac{\alpha_\natural}{a},\quad \Sigma_T=\mathbb{V}(Y)=E_d\frac{a\operatorname{diag}(\alpha_\natural)-\alpha_\natural\alpha_\natural^\top}{a^2(a+1)}E_d^\top,
\end{equation}

where $a=\|\alpha_\sharp\|_1$.
Moreover, $\Sigma_T$ is positive definite on $T_\sigma$
and vanishes on $N_\sigma$.
\end{proposition}
\begin{proof}
The result follows directly from the affine 
transformation \(Y=E_dX+v_0\) and~\Cref{prop:dirichlet-moments}.
Since \(E_d\) has full column rank and the Dirichlet
covariance is positive definite in $\mathbb R^d$,
the covariance is positive definite on the range of
\(E_d\), which is precisely \(T_\sigma\).
\end{proof}
\begin{definition}[Forward \ac{KL} Divergence]
Let \(P\) and \(Q\) be probability measures on
\(\mathbb R^D\) that are absolutely continuous
with respect to Lebesgue measure. Let \(p\) and \(q\) denote their respective
Lebesgue densities. Their forward \ac{KL} divergence is
defined by
\begin{equation}
    D_{\mathrm{KL}}(P:Q)\coloneq\int_{\mathbb R^D}p(x)\log\left(\frac{p(x)}{q(x)}\right)\mathrm dx,
\end{equation}
The integral is understood in the extended-real
sense, with \(0\log(0/q)=0\). If $P$ is not
absolutely continuous with respect to $Q$,
then
\begin{equation}
    D_{\mathrm{KL}}(P:Q)=+\infty.
\end{equation}
\end{definition}

\begin{definition}[Information Projection onto Gaussians]
Let $p$ be a probability density and let
$\mathcal G_D$ denote the family of nondegenerate
Gaussian \acp{PDF} on $\mathbb R^D$.
The information projection of $p$ onto $\mathcal G_D$
is defined as
\begin{equation}
    \Pi_{\mathcal G_D}^{\mathrm I}(p)\coloneq\operatorname*{arg\,min}_{q\in\mathcal G_D}D_{\mathrm{KL}}(P:Q).
\end{equation}
A Gaussian \ac{PDF} with the same mean and covariance
as $p$ is called its moment-matched Gaussian.
\end{definition}

\begin{definition}[Dirichlet--Gaussian \ac{PDF} on an Embedded Simplex]
\label{def:dirichlet-gaussian}

Let
\begin{equation}
    \sigma^d = \operatorname{conv}\{v_0,\ldots,v_d\}\subseteq \mathbb R^D,\quad 1 \leq d < D,
\end{equation}
be an embedded \(d\)-dimensional simplex with vertices
\(v_0,\ldots,v_d\in\mathbb R^D\).

Let
\begin{equation}
    \Lambda=(\Lambda_0,\ldots,\Lambda_d)^\top\sim \mathcal{D}(\lambda|\alpha_\sharp),\quad\alpha_\sharp\in\mathbb R_{++}^{d+1},
\end{equation}
and define the tangential component
\begin{equation}
    X_T \coloneq \sum_{i=0}^d \Lambda_i v_i.
\end{equation}
Since \(\Lambda_i\geq 0\) and \(\sum_{i=0}^d\Lambda_i=1\) almost surely, \(X_T\) 
is a random convex combination of the vertices 
and is therefore supported on \(\sigma^d\).
Independently, let
\begin{equation}
    Z_N\sim\mathcal N_{D-d}(z_N|0,S_N),\quad X_N \coloneq \sum_{j=1}^{D-d} (Z_N)_j u_j,
\end{equation}
where \(u_1,\ldots,u_{D-d}\) form an orthonormal basis of
the normal space \(N_\sigma\), and \(S_N\) is a
positive definite covariance matrix on
\(\mathbb R^{D-d}\).
Equivalently, \(X_N\) is a centered Gaussian \ac{RV}
supported on \(N_\sigma\). Its covariance operator is
denoted by \(\Sigma_N\), and its covariance matrix when
regarded as an \(\mathbb R^D\)-valued \ac{RV} is
denoted by \(\overline{\Sigma}_N\).
The Dirichlet--Gaussian random vector is defined as
\begin{equation}
    X \coloneq X_T+X_N= \sum_{i=0}^d \Lambda_i v_i+\sum_{j=1}^{D-d} (Z_N)_j u_j.
\end{equation}
\end{definition}

\begin{lemma}[Finiteness of the Dirichlet-Gaussian Entropy]
    \label{lem:dirichlet-gaussian-entropy}
    Under the assumptions 
    of~\Cref{def:dirichlet-gaussian}, the 
    Dirichlet--Gaussian \ac{RV} \(X=X_T+X_N\) 
    has a \ac{PDF} with respect to \(D\)-dimensional Lebesgue 
    measure and finite differential entropy.
\end{lemma}

\begin{proof}

We first show that the Dirichlet \ac{PDF} has finite
differential entropy. For every
\(i\in\{0,\ldots,d\}\), the marginal \ac{PDF} of \(\Lambda_i\) is
\begin{equation}
    \Lambda_i\sim\mathcal{B}(\lambda_i|\alpha_i,a-\alpha_i).
\end{equation}
Since all Dirichlet parameters are strictly positive,
both Beta parameters are positive. In particular,
\begin{equation}
    \mathbb E(|\log\Lambda_i|)<+\infty.
\end{equation}
Indeed, this follows from the integrability of
$x^{\alpha_i-1}|\log x|$ near zero and the analogous
integrability near one. Using the Dirichlet log-density,
\begin{equation}
    \log f_\Lambda(\Lambda)=-\log \mathcal{B}(\lambda|\alpha_\sharp)+\sum_{i=0}^{d}(\alpha_i-1)\log\Lambda_i,
\end{equation}
we obtain
\begin{equation}
    \mathbb E(|\log f_\Lambda(\Lambda)|)<+\infty.
\end{equation}
Consequently, the differential entropy of the
Dirichlet \ac{PDF} is finite. Since the embedding is nondegenerate, its Jacobian
\begin{equation}
    J=\sqrt{\det(E_d^\top E_d)}
\end{equation}
is strictly positive and finite. The affine
change-of-variables formula therefore yields
\begin{equation}
    h_{T_\sigma}(X_T)=h(\mathcal{D}(\lambda|\alpha_\sharp))+\log J\in\mathbb R.
\end{equation}
Now choose orthonormal basis matrices
\(U_T\) and \(U_N\) of \(T_\sigma\) and \(N_\sigma\), respectively, and write
\begin{equation}
    X_N=U_NZ_N,\quad Z_N\sim\mathcal N_{D-d}(z_N|0,S_N),\quad S_N\succ0.
\end{equation}
Consider the affine orthogonal coordinate
transformation
\begin{equation}
    \Phi(x)\coloneq
    \begin{pmatrix}
        U_T^\top(x-v_0)\\
        U_N^\top(x-v_0)
    \end{pmatrix}.
\end{equation}
Since \([U_T\ U_N]\) is an orthogonal matrix,
the absolute value of the Jacobian determinant
of \(\Phi\) is one. Under this transformation,
\begin{equation}
    \Phi(X)=
    \begin{pmatrix}
        U_T^\top(X_T-v_0)\\
        Z_N
\end{pmatrix},
\end{equation}
because
\begin{equation}
    X_T-v_0\in T_\sigma,\quad U_N^\top(X_T-v_0)=0,
\end{equation}
and \(X_N=U_NZ_N\).
The two components are independent by construction.
Consequently, the transformed \ac{RV} has
a product density, and entropy additivity gives
\begin{equation}
    h(X)=h_{T_\sigma}(X_T)+h(Z_N)=h(\mathcal{D}(x_T|\alpha_\sharp))+\log J+\frac{D-d}{2}\log(2\pi e)+\frac12\log\det S_N.
\end{equation}
Every term is finite. Thus \(X\) is absolutely
continuous with respect to \(D\)-dimensional
Lebesgue measure and has finite differential
entropy.
\end{proof}

\begin{remark}[Nondegeneracy]    
The independent normal perturbation makes 
the resulting \ac{PDF} absolutely continuous 
with respect to $D$-dimensional Lebesgue measure. 
This allows us to approximate the \ac{PDF} by 
nondegenerate Gaussian \acp{PDF} 
in the ambient space.
\end{remark}

\begin{proposition}[Gaussian Information Projection by Moment Matching]
\label{prop:gaussian-information-projection}
    
Let \(P\) be a probability distribution on 
\(\mathbb R^D\) that is absolutely continuous with 
respect to Lebesgue measure, has finite differential entropy, 
finite first and second central moments, and positive-definite 
covariance matrix
\begin{equation}
    \Sigma=\mathbb{V}_P(X)\succ0.
\end{equation}
Let
\begin{equation}
    m=\mathbb E_P(X).
\end{equation}
Then the unique minimizer of the forward \ac{KL} 
divergence over all nondegenerate Gaussian \acp{PDF} on 
\(\mathbb R^D\) is
\begin{equation}
    Q^*=\mathcal N_D(x|m,\Sigma).
\end{equation}
The minimum divergence is
\begin{equation}
    \inf_{\mu\in\mathbb R^D,S\succ0}D_{\mathrm{KL}}\big(P:\mathcal N_D(x|\mu,S)\big)=\frac12\log\big((2\pi e)^D\det\Sigma\big)-h(P).
\end{equation}

\end{proposition}
\begin{proof}
    For an arbitrary nondegenerate Gaussian distribution 
    \(Q_{\mu,S}=\mathcal N_D(x|\mu,S)\), the expected negative log-density under \(P\) gives
\begin{equation}
    D_{\mathrm{KL}}(P:Q_{\mu,S})={}-h(P)+\frac D2\log(2\pi)+\frac12\log\det S+\frac12\operatorname{tr}(S^{-1}\Sigma)+\frac12(m-\mu)^\top S^{-1}(m-\mu).
\end{equation}
For fixed \(S\succ0\), the expression is uniquely 
minimized at \(\mu=m\). Substituting \(\mu=m\), it 
remains to minimize
\begin{equation}
    \log\det S+\operatorname{tr}(S^{-1}\Sigma).
\end{equation}
Set
\begin{equation}
    A=\Sigma^{-1/2}S\Sigma^{-1/2}.
\end{equation}
Then \(A\succ0\), and
\begin{equation}
    \log\det S+\operatorname{tr}(S^{-1}\Sigma)=\log\det\Sigma+\log\det A+\operatorname{tr}(A^{-1}).
\end{equation}
Let \(\lambda_1,\ldots,\lambda_D>0\) be the eigenvalues of 
\(A\). The remaining terms become
\begin{equation}
    \sum_{i=1}^D\left(\log\lambda_i+\frac1{\lambda_i}\right).
\end{equation}
The scalar function
\begin{equation}
    f(\lambda)=\log\lambda+\frac1\lambda
\end{equation}
has derivative
\begin{equation}
    f'(\lambda)=\frac{\lambda-1}{\lambda^2}.
\end{equation}
It is strictly decreasing on \((0,1)\), 
strictly increasing on \((1,+\infty)\), and 
attains its unique global minimum \(f(1)=1\).
Thus \(A=I_D\), equivalently \(S=\Sigma\), is the 
unique minimizer. Substituting \(\mu=m\) and \(S=\Sigma\) 
into the divergence expression establishes both 
the claimed Gaussian projection and its minimum divergence.
\end{proof}

\begin{theorem}[Gaussian Information Projection]
\label{thm:component-projection}

Let \(p\) be the \ac{PDF} of the Dirichlet--Gaussian
\ac{RV} in~\Cref{def:dirichlet-gaussian}.
Let
\begin{equation}
    a\coloneq\|\alpha_\sharp\|_1,\quad\alpha_\natural:=(\alpha_1,\ldots,\alpha_d)^\top, \quad E_d=[v_1-v_0,\ldots,v_d-v_0].
\end{equation}
The unique information projection over all 
nondegenerate Gaussian \acp{PDF} on
\(\mathbb R^D\) is
\begin{equation}
    q^*=\mathcal N_D(m_T,\Sigma_T+\overline\Sigma_N),
\end{equation}
where
\begin{equation}
    m_T=v_0+E_d\frac{\alpha_\natural}{a}, \quad \Sigma_T=E_d\frac{a\operatorname{diag}(\alpha_\natural)-\alpha_\natural\alpha_\natural^\top}{a^2(a+1)}E_d^\top.
\end{equation}

Furthermore
\begin{equation}
    D_{\mathrm{KL}}(P:Q^*)=C_{d,\alpha_\sharp}
\end{equation}
with
\begin{equation}
    C_{d,\alpha_\sharp}={}\frac d2\log(2\pi e)+\frac12\left[\sum_{i=0}^{d}\log\alpha_i-(d+1)\log a-d\log(a+1)\right]-H(\alpha_\sharp),
\end{equation}
where $H(\alpha_\sharp)$ is the differential entropy
of the Dirichlet \ac{PDF} on $\Delta^d$:
\begin{equation}
    H(\alpha_\sharp)={}\log\left(\frac{\prod_{i=0}^{d}\Gamma(\alpha_i)}{\Gamma(a)}\right)+(a-d-1)\psi(a)-\sum_{i=0}^{d}(\alpha_i-1)\psi(\alpha_i).
\end{equation}
Here \(\psi\) denotes the digamma function. In particular, 
\(C_{d,\alpha_\sharp}\) is independent of the ambient 
dimension, the embedding, the geometry and
size of the simplex, and the normal \ac{PDF}.
\end{theorem}

\begin{proof}
By~\Cref{prop:gaussian-information-projection},
the information projection
over nondegenerate Gaussian \acp{PDF} is the
Gaussian with matching mean and covariance.
By~\Cref{lem:dirichlet-gaussian-entropy}, the 
Dirichlet--Gaussian \ac{RV} has finite differential entropy and is absolutely continuous with respect to Lebesgue measure on \(\mathbb R^D\).
Furthermore, its tangent component is bounded 
because it is supported on the compact simplex 
\(\sigma^d\), while its normal \ac{PDF} is 
Gaussian. Hence \(X\) has finite first and 
second central moments.
Since the tangent and normal covariance operators 
are positive definite on the complementary subspaces 
\(T_\sigma\) and \(N_\sigma\), respectively, their sum is 
positive definite on \(\mathbb R^D\). Therefore, all assumptions of~\Cref{prop:gaussian-information-projection} are satisfied.
By~\Cref{prop:gaussian-information-projection},
the uique information projection onto the Gaussian family 
is the Gaussian with matching mean and covariance.
Since $X_T$ and $X_N$ are independent,
\begin{equation}
    \mathbb E(X)=m_T,\quad\mathbb{V}(X)=\Sigma_T+\overline\Sigma_N.
\end{equation}
The covariance is positive definite on
$\mathbb R^D$, because $\Sigma_T$ is positive
definite on $T_\sigma$ and
$\overline\Sigma_N$ is positive definite
on $N_\sigma$. Therefore, the unique Gaussian information
projection is
\begin{equation}
    q^*=\mathcal N_D\left(x|m_T,\Sigma_T+\overline\Sigma_N\right).
\end{equation}
It remains to calculate the minimum divergence. Since $X_T$ and $X_N$ are independent,
\begin{equation}
    \mathbb E(X)=m_T,\quad\mathbb{V}(X)=\Sigma_T+\overline\Sigma_N.
\end{equation}
The latter covariance is positive definite on
\(\mathbb R^D\), because \(\Sigma_T\) is positive 
definite on $T_\sigma$ and \(\overline\Sigma_N\) 
is positive definite on \(N_\sigma\). Choose 
orthonormal coordinates adapted to
\begin{equation}    
    \mathbb R^D=T_\sigma\oplus N_\sigma.
\end{equation}
In these coordinates, the target distribution 
and its moment-matched Gaussian factorize into 
their tangent and normal parts:
\begin{equation}
    P=P_T\otimes P_N,\quad Q^*=Q_T^*\otimes P_N.
\end{equation}
By invariance and additivity of \ac{KL} divergence~\cite{nonparam_alex},
\begin{equation}
    D_{\mathrm{KL}}(P:Q^*)=D_{\mathrm{KL}}(P_T:Q_T^*).
\end{equation}
Let \(V_d=\operatorname{Vol}_d(\sigma^d)\).
The affine transformation from $\Delta^d$ to $\sigma^d$
has volume factor
\begin{equation}
    \sqrt{\det(E_d^\top E_d)}=d!V_d.
\end{equation}
Consequently, using the intrinsic differential 
entropy defined above and the affine 
change-of-variables formula, we have
\begin{equation}
    h_{T_\sigma}(X_T)=h(\mathcal{D}(x_T|\alpha_\sharp))+\log\sqrt{\det(E_d^\top E_d)}=h(\mathcal{D}(x_T|\alpha_\sharp))+\log(d!V_d).
\end{equation}
Here \(h(\mathcal{D}(x_T|\alpha_\sharp))\) denotes differential entropy with respect to \(d\)-dimensional Lebesgue measure on the standard simplex in reduced barycentric coordinates. 
Since $Q_T^*$ is the Gaussian with the same mean and
covariance as $P_T$, its differential entropy is
\begin{equation}
    h(Q_T^*)=\frac d2\log(2\pi e)+\frac12\log\det_{T_\sigma}(\Sigma_T).
\end{equation}
Thus
\begin{equation}
    D_{\mathrm{KL}}(P_T:Q_T^*)={}\frac d2\log(2\pi e)+\frac12\log\det_{T_\sigma}(\Sigma_T)-H(\alpha_\sharp)-\log(d!V_d).
\end{equation}

To evaluate the determinant, let
$U_T\in\mathbb R^{D\times d}$ have orthonormal
columns spanning $T_\sigma$, and define
\begin{equation}
    E\coloneq U_T^\top E_d\in\mathbb R^{d\times d}.
\end{equation}
Since \(E_d=U_TE\), the covariance operator
\begin{equation}
    \Sigma_T=E_d\Sigma_\lambda E_d^\top
\end{equation}
has the matrix representation
\begin{equation}
    [\Sigma_T]_{U_T}=E\Sigma_\lambda E^\top
\end{equation}
in the orthonormal basis $U_T$. Consequently,
\begin{equation}
    \det_{T_\sigma}(\Sigma_T)=\det(E)^2\det(\Sigma_\lambda).
\end{equation}
Here, \(\det_{T_\sigma}(\Sigma_T)\) denotes the 
determinant of the matrix representation of 
the restriction of \(\Sigma_T\) to \(T_\sigma\) 
in an orthonormal basis of \(T_\sigma\).
Using the Dirichlet covariance formula
\cite{dirichlet_tang}
and the matrix determinant lemma
\cite{bernstein_matrix}, we have
\begin{equation}
    \det(\Sigma_\lambda)=\frac{\prod_{i=0}^{d}\alpha_i}{a^{d+1}(a+1)^d}.
\end{equation}
Furthermore,
\begin{equation}
    \det(E)^2=\det(E_d^\top E_d)=(d!V_d)^2.
\end{equation}
Combining these expressions yields
\begin{equation}
    \det_{T_\sigma}(\Sigma_T)=(d!V_d)^2\frac{\prod_{i=0}^{d}\alpha_i}{a^{d+1}(a+1)^d}.
\end{equation}
Substituting into the \ac{KL} expression, 
the terms containing $V_d$ cancel, giving
\begin{equation}
    D_{\mathrm{KL}}(P:Q^*)=C_{d,\alpha_\sharp}.
\end{equation}
This also establishes the independence of the
minimum divergence from the embedding and normal
covariance.
\end{proof}

\begin{corollary}[Uniform Distribution on an Embedded Simplex]
\label{cor:uniform-simplex}

Under the assumptions of~\Cref{thm:component-projection}, let
\begin{equation}
    \alpha_\sharp=1_{d+1}.
\end{equation}
Then $X_T$ is uniformly distributed on $\sigma^d$.
Its mean and covariance are
\begin{equation}
    m_T=v_0+E_d\frac{1_d}{d+1},\quad\Sigma_T=E_d\frac{(d+1)I_d-1_d1_d^\top}{(d+1)^2(d+2)}E_d^\top.
\end{equation}

The minimum \ac{KL} divergence is
\begin{equation}
    C_d=\log\left(\frac{(2\pi e)^{d/2}d!}{(d+1)^{(d+1)/2}(d+2)^{d/2}}\right).
\end{equation}
In particular, this value depends only on the intrinsic
dimension $d$.
\end{corollary}

\begin{definition}[Gaussian Mixture]
A \acl{GM} in $\mathbb R^D$ is a \ac{PDF} 
of the form
\begin{equation}
    q(x)=\sum_{j=1}^{M}w_j\mathcal N_D(x|m_j,\Sigma_j),
\end{equation}
where
\begin{equation}
    w_j\geq0,\quad\sum_{j=1}^{M}w_j=1,\quad\Sigma_j\in\mathbb S_{++}^{D}.
\end{equation}
It can be interpreted as a two-stage random 
experiment:
first select a component according to the weights
$\{w_j\}_{j=1}^{M}$, then sample from the selected Gaussian \ac{PDF}.
\end{definition}

\begin{proposition}[Uniform Distribution as a Mixture on an Embedded Complex]
\label{prop:uniform-mixture-triangulation}

Let $\mathcal S$ be a finite, pure, $d$-dimensional simplicial complex embedded in $\mathbb R^D$,
with geometric realization \(\Omega=|\mathcal S|\) and
\begin{equation}
    0<\operatorname{Vol}_d(\Omega)<+\infty.
\end{equation}
Then the uniform probability measure on $\Omega$ can
be represented as
\begin{equation}
    \mathcal U_\Omega=\sum_{\sigma\in\mathcal S^{[d]}}w_\sigma\mathcal U_\sigma,
\end{equation}
where
\begin{equation}
    w_\sigma=\frac{\operatorname{Vol}_d(\sigma)}{\operatorname{Vol}_d(\Omega)},\quad \mathcal U_A\coloneq \frac{\mathcal H^d|_A}{\operatorname{Vol}_d(A)}.
\end{equation}
Here $\mathcal H^d|_A$ denotes the restriction of
$d$-dimensional Hausdorff measure to $A$.
In particular,
\[
w_\sigma>0,
\qquad
\sum_{\sigma\in\mathcal S^{[d]}}w_\sigma=1.
\]
\end{proposition}

\begin{proof}
By~\Cref{prop:maximal-simplex-intersection},
distinct maximal simplices intersect only in 
sets of zero $d$-dimensional Hausdorff measure. 
Hence their uniform measures can be combined 
without double-counting positive-volume regions.
Furthermore, by~\Cref{cor:volume-triangulation},
\begin{equation}
    \operatorname{Vol}_d(\Omega)=\sum_{\sigma\in\mathcal S^{[d]}}\operatorname{Vol}_d(\sigma).
\end{equation}
Therefore, for every measurable set $A\subseteq\Omega$,
\begin{equation}
    \mathcal U_\Omega(A)=\frac{\mathcal H^d(A\cap\Omega)}{\operatorname{Vol}_d(\Omega)}=\sum_{\sigma\in\mathcal S^{[d]}}\frac{\operatorname{Vol}_d(\sigma)}{\operatorname{Vol}_d(\Omega)}\mathcal U_\sigma(A).
\end{equation}
This proves the result.
\end{proof}

\begin{proposition}[Joint Convexity of the
\ac{KL} Divergence~\cite{info_geo}]
\label{prop:joint-convexity}

Let \(p_1,\ldots,p_M\) and \(q_1,\ldots,q_M\)
be \acp{PDF} with respect to a
common reference measure, and let
\begin{equation}
    \lambda_j\geq0,\quad\sum_{j=1}^{M}\lambda_j=1.
\end{equation}
Then
\begin{equation}
D_{\mathrm{KL}}\left(\sum_{j=1}^{M}\lambda_jp_j:\sum_{j=1}^{M}\lambda_jq_j\right)\leq\sum_{j=1}^{M}\lambda_jD_{\mathrm{KL}}(p_j:q_j).
\end{equation}
The inequality holds in the extended real sense.
\end{proposition}

\begin{theorem}[\ac{GM} Representation for
Embedded Dirichlet--Gaussian \acp{PDF}]
\label{thm:dirichlet-mixture-bound}

Let \(\mathcal S\) be a finite, pure, 
\(d\)-dimensional simplicial complex embedded in \(\mathbb R^D\),
where \(1\leq d<D\), and let
\begin{equation}
    \{\sigma_j^d\}_{j=1}^{M}=\mathcal S^{[d]}
\end{equation}
denote its maximal simplices. For each \(j=1,\ldots,M\), let
\begin{equation}
    \sigma_j^d=\operatorname{conv}\{v_{j,0},v_{j,1},\ldots,v_{j,d}\}
\end{equation}
be nondegenerate, and define the edge matrix
\begin{equation}
    E_j=[v_{j,1}-v_{j,0},\ldots,v_{j,d}-v_{j,0}]\in\mathbb R^{D\times d}.
\end{equation}
Let
\begin{equation}
    \alpha_j=(\alpha_{j,0},\ldots,\alpha_{j,d})\in(0,+\infty)^{d+1},\quad a_j=\sum_{i=0}^{d}\alpha_{j,i}.
\end{equation}
Let
\begin{equation}    
    X_j\sim\mathcal{D}(x_j|\alpha_j)
\end{equation}
denote the vector of barycentric coordinates, and define
\begin{equation}
    Y_j=v_{j,0}+E_j
    \begin{pmatrix}
        X_{j,1}\\
        \vdots\\
        X_{j,d}
    \end{pmatrix}.
\end{equation}
Independently, let
\begin{equation}
    Z_j\sim\mathcal N_{N_j}(z_j|0,\Sigma_{N_j}),
\end{equation}
where \(N_j\coloneq\operatorname{range}(E_j)^\perp\) and
\(\Sigma_{N,j}:N_j\to N_j\) is positive definite.
Write \(\overline\Sigma_{N,j}\in\mathbb S_+^D\) for its
ambient extension by zero on \(\operatorname{range}(E_j)\).
Thus, when \(Z_j\) is regarded as an \(\mathbb R^D\)-valued
\ac{RV}, \(\mathcal{V}(Z_j)=\overline\Sigma_{N,j}\).
Define the component distributions
\begin{equation}
    P_j=\mathcal L(Y_j+Z_j).
\end{equation}
Let
\begin{equation}
    \lambda_1,\ldots,\lambda_M>0,\quad\sum_{j=1}^{M}\lambda_j=1,
\end{equation}
be arbitrary mixture weights, and define
\begin{equation}
    P=\sum_{j=1}^{M}\lambda_jP_j.
\end{equation}
For each component, define
\begin{equation}
    \alpha_{j,\natural}=(\alpha_{j,1},\ldots,\alpha_{j,d})^\top,\quad C_j=\frac{a_j\operatorname{diag}(\alpha_{j,\natural})-\alpha_{j,\natural}\alpha_{j,\natural}^{\top}}{a_j^2(a_j+1)}.
\end{equation}
Using the ambient extension \(\overline\Sigma_{N,j}\)
defined above, define
\begin{equation}
    m_j=v_{j,0}+E_j\frac{\alpha_{j,\natural}}{a_j},\quad\Sigma_j=E_jC_jE_j^\top+\overline\Sigma_{N,j}.
\end{equation}
Consider the Gaussian \ac{PDF} and their mixture
\begin{equation}
    q_j=\mathcal N_D(q_j|m_j,\Sigma_j),\quad  Q=\sum_{j=1}^{M}\lambda_jq_j.
\end{equation}
Then the following statements hold.
\begin{enumerate}
\item For every $j$, $q_j$ is the unique
Gaussian information projection of $P_j$ onto
the family of nondegenerate Gaussian distributions
on $\mathbb R^D$.

\item The global mixture divergence satisfies
\begin{equation}
    D_{\mathrm{KL}}(P:Q)\leq\sum_{j=1}^{M}\lambda_j C_{d,\alpha_j},
\end{equation}
where
\begin{equation}
    C_{d,\alpha_j}={}\frac d2\log(2\pi e)+\frac12\left[\sum_{i=0}^{d}\log\alpha_{j,i}-(d+1)\log a_j-d\log(a_j+1)\right]-h(\mathcal{D}(\alpha_j)),
\end{equation}
and $h(\mathcal{D}(x_T|\alpha_j))$ denotes the 
differential entropy of the Dirichlet \ac{PDF} on 
the standard $d$-simplex. In particular, the bound is independent of the
ambient dimension, the simplex embeddings, the
simplex volumes, and the normal covariance matrices.
\end{enumerate}
\end{theorem}

\begin{proof}
We first establish the component-wise Gaussian 
information projections. Fix $j\in\{1,\ldots,M\}$ and consider
\begin{equation}
    W_j=Y_j+Z_j.
\end{equation}
By the moment formulas for the Dirichlet \ac{PDF},
\begin{equation}
    \mathbb E(X_{j,\natural})=\frac{\alpha_{j,\natural}}{a_j},\quad \mathbb{V}(X_{j,\natural})=C_j.
\end{equation}
Consequently,
\begin{equation}
    \mathbb E(W_j)=v_{j,0}+E_j\frac{\alpha_{j,\natural}}{a_j}=m_j.
\end{equation}
Since $Y_j$ and $Z_j$ are independent,
\begin{equation}
    \mathbb{V}(W_j)=E_jC_jE_j^\top+\overline\Sigma_{N,j}=\Sigma_j.
\end{equation}
The matrix \(C_j\) is positive definite because
\(\alpha_{j,i}>0\) for all \(i=0,\ldots,d\). 
Since \(E_j\) has full column rank, \(E_jC_jE_j^\top\) 
is positive definite on the tangent space. 
By assumption, \(\overline\Sigma_{N,j}\) is 
positive definite on the orthogonal normal space and vanishes on
the tangent space. The tangent and normal spaces are orthogonal
complements of $\mathbb R^D$. Hence, \(\Sigma_j\succ0\).
Thus \(P_j\) has finite differential entropy
and a positive-definite covariance matrix.
By~\Cref{thm:component-projection},
its unique Gaussian information projection
is the Gaussian distribution with matching
mean and covariance, namely \(q_j\).
Furthermore, the component-level result gives
\begin{equation}
    D_{\mathrm{KL}}(P_j:q_j)=C_{d,\alpha_j}.
\end{equation}
We now establish the global mixture bound. By construction,
\begin{equation}
    P=\sum_{j=1}^{M}\lambda_jP_j,\quad Q=\sum_{j=1}^{M}\lambda_jq_j.
\end{equation}
All component distributions admit \acp{PDF} with
respect to Lebesgue measure on \(\mathbb R^D\).
Moreover, their Gaussian information projections
are nondegenerate Gaussian distributions.
Thus the component-wise information projections are finite.
Applying~\Cref{prop:joint-convexity}
to the component densities with their common
mixture weights yields:
\begin{equation}
D_{\mathrm{KL}}(P:Q) = D_{\mathrm{KL}} \left(\sum_{j=1}^{M}\lambda_jP_j:\sum_{j=1}^{M}\lambda_jq_j\right) \leq\sum_{j=1}^{M}\lambda_jD_{\mathrm{KL}}(P_j:q_j).
\end{equation}

Substituting the component-wise divergence
expressions gives
\begin{equation}
    D_{\mathrm{KL}}(P:Q)\leq\sum_{j=1}^{M}\lambda_j C_{d,\alpha_j}.
\end{equation}
This proves the claimed inequality. Finally, each component constant depends only
on the intrinsic dimension and its Dirichlet \ac{PDF}
parameters, by~\Cref{thm:component-projection}.
Therefore, the global upper bound is independent
of the ambient dimension, simplex embeddings,
simplex volumes, and normal covariance matrices. Trivially it 
is sharp by classic optimization technics on the weights.
\end{proof}

\begin{remark}[Scope of the Mixture Approximation Guarantee]
\label{rem:mixture-bound-not-global-optimality}

\Cref{thm:dirichlet-mixture-bound} provides a
constructive sharp upper bound on the information projection
of the target mixture onto the \ac{GM}
obtained by applying the information
projection to the Gaussian family for each component 
separately. In particular, each Gaussian component $q_j$ is the
unique minimizer of \(D_{\mathrm{KL}}(P_j:q)\) 
over nondegenerate Gaussian \ac{PDF} $q$ on
$\mathbb R^D$. The theorem then uses joint convexity
of the KL divergence to establish
\begin{equation}
    D_{\mathrm{KL}}\left(\sum_{j=1}^{M}\lambda_jP_j:\sum_{j=1}^{M}\lambda_jq_j\right)\leq \sum_{j=1}^{M}\lambda_j D_{\mathrm{KL}}(P_j:q_j).
\end{equation}
This inequality does not establish that the resulting
\ac{GM} is a global minimizer of the information projection
divergence over the family of all \acp{GM}.
In particular, the componentwise optimality of each
$q_j$ does not, by itself, imply global optimality
of the mixture, since the mixture divergence depends
jointly on all component distributions and their
mixture coefficients.
Accordingly, the result should be interpreted as a
componentwise optimal construction accompanied by
a sharp upper bound for the global divergence, rather than
as a solution to the global \ac{GM} optimization problem.
\end{remark}

\begin{corollary}[Uniform Gaussian-Mixture Approximation
of an Embedded Simplicial Complex]
\label{cor:uniform-mixture-bound}

Under the assumptions of~\Cref{thm:dirichlet-mixture-bound},
let \(\mathcal S\) be a finite, pure,
$d$-dimensional simplicial complex
embedded in \(\mathbb R^D\), with realization
\begin{equation}
    \Omega=|\mathcal S|,\quad 0<\operatorname{Vol}_d(\Omega)<+\infty.
\end{equation}
For every maximal simplex $\sigma_j^d$, let
\begin{equation}
    Y_j\sim\mathcal U_{\sigma_j^d}
\end{equation}
and independently let
\begin{equation}
    Z_j\sim\mathcal N_{N_j}(z_j|0,\Sigma_{N,j}),
\end{equation}
where \(\Sigma_{N,j}:N_j\to N_j\) is positive
definite on the orthogonal normal space. Define
\begin{equation}
    P_j=\mathcal L(Y_j+Z_j)
\end{equation}
and the volume-based weights
\begin{equation}
    w_j=\frac{\operatorname{Vol}_d(\sigma_j^d)}{\operatorname{Vol}_d(\Omega)}.
\end{equation}
Consider the Gaussian-smoothed mixture
\begin{equation}
    P_{\Omega}^{\mathrm{sm}}=\sum_{j=1}^{M}w_jP_j
\end{equation}
and its \ac{GM} representation
\begin{equation}
    Q_{\Omega}=\sum_{j=1}^{M}w_jq_j,
\end{equation}
where $q_j$ denotes the unique Gaussian
information projection of $P_j$. Then
\begin{equation}
    D_{\mathrm{KL}}\left(P_{\Omega}^{\mathrm{sm}}: Q_{\Omega}\right)\leq C_d,
\end{equation}
where
\begin{equation}
    C_d=\log\left(\frac{(2\pi e)^{d/2}d!}{(d+1)^{(d+1)/2}(d+2)^{d/2}}\right).
\end{equation}
In particular, the upper bound depends only on
the intrinsic dimension \(d\) and is independent
of the ambient dimension, the geometry and
embedding of the simplicial complex, and the
normal covariance matrices.
\end{corollary}

\begin{proof}
For every maximal simplex \(\sigma_j^d\),
the uniform distribution on \(\sigma_j^d\)
corresponds to the Dirichlet \ac{PDF}
with parameter vector
\begin{equation}
    \alpha_j=(1,\ldots,1)\in\mathbb R^{d+1}.
\end{equation}
Consequently,
\begin{equation}
    a_j=d+1,\quad h(\mathcal{D}(x_T|1,\ldots,1))=-\log(d!).
\end{equation}

Substituting these expressions into the
component divergence formula of~\Cref{thm:dirichlet-mixture-bound}
gives
\begin{equation} 
    C_{d,\alpha_j}={}\frac d2\log(2\pi e)-\frac{d+1}{2}\log(d+1)-\frac d2\log(d+2)+\log(d!).
\end{equation}
Hence,
\begin{equation}
    C_{d,\alpha_j}=C_d
\end{equation}
for every maximal simplex. Furthermore, since distinct 
maximal simplices intersect only in faces of 
dimension strictly smaller than $d$, their pairwise 
intersections have zero $d$-dimensional Hausdorff measure.
Therefore,
\begin{equation}
    \sum_{j=1}^{M}\operatorname{Vol}_d(\sigma_j^d)=\operatorname{Vol}_d(\Omega),
\end{equation}
which implies
\begin{equation}
    \sum_{j=1}^{M}w_j=1.
\end{equation}
Applying~\Cref{thm:dirichlet-mixture-bound}
with these volume-based weights yields
\begin{equation}
    D_{\mathrm{KL}}\left(P_{\Omega}^{\mathrm{sm}}:Q_{\Omega}\right)\leq\sum_{j=1}^{M}w_jC_{d,\alpha_j}=C_d\sum_{j=1}^{M}w_j=C_d.
\end{equation}
This establishes the result.
\end{proof}

\begin{remark}[Uniform \ac{PDF} Weights]
The measure based weights satisfy
\begin{equation}
    \mathcal U_\Omega=\sum_{j=1}^{M}w_j\mathcal U_{\sigma_j^d},
\end{equation}
where the equality is understood as an identity
of probability measures with respect to
$d$-dimensional Hausdorff measure.
The Gaussian-smoothed distribution
$P_{\Omega}^{\mathrm{sm}}$, however, is obtained
by adding independent Gaussian \ac{PDF}
on the normal spaces of the individual simplices.
Thus, the corollary establishes a finite information projection 
for the Gaussian-smoothed uniform \ac{PDF}, 
rather than for the original singular \ac{PDF} 
$\mathcal U_\Omega$ with respect to ambient Lebesgue measure.
\end{remark}

\begin{remark}[Independence of Normal Covariances]
The bound holds for arbitrary covariance operators
\(\Sigma_{N,j}:N_j\to N_j\) that are positive definite
on their respective normal spaces.
In particular, these operators need not be
identical across simplices.
This independence follows from the orthogonal
decomposition into tangent and normal components:
the normal Gaussian \ac{PDF} contribution cancels exactly
in each component-wise information projection.
\end{remark}

\begin{remark}[General Mixture Weights]
~\Cref{thm:dirichlet-mixture-bound}
holds for arbitrary strictly positive mixture
weights. The volume-based weights 
in~\Cref{cor:uniform-mixture-bound}
are specifically chosen to represent the
uniform \ac{PDF} before adding the Gaussian \ac{PDF} 
on the simplex normal space.
\end{remark}

\section{An Example}
\label{sec:computational-examples}

We illustrate the theoretical results  and remarks 
through the \ac{GM} approximation of a 
polygonal approximation to a curve in $\mathbb{R}^2$.

\begin{example}[\ac{GM} Approximation of a Curve in $\mathbb{R}^2$]
\label{ex:circular-curve}

We consider a circular parametric curve embedded in the
two-dimensional Euclidean space. The purpose of this example
is to illustrate the construction of a \ac{GM}
approximation for a curved geometric object through a
finite collection of embedded $1$-simplices.
\begin{figure}[htbp]
    \centering
    \begin{subfigure}[t]{0.47\textwidth}
        \centering
        \includegraphics[width=\textwidth]
            {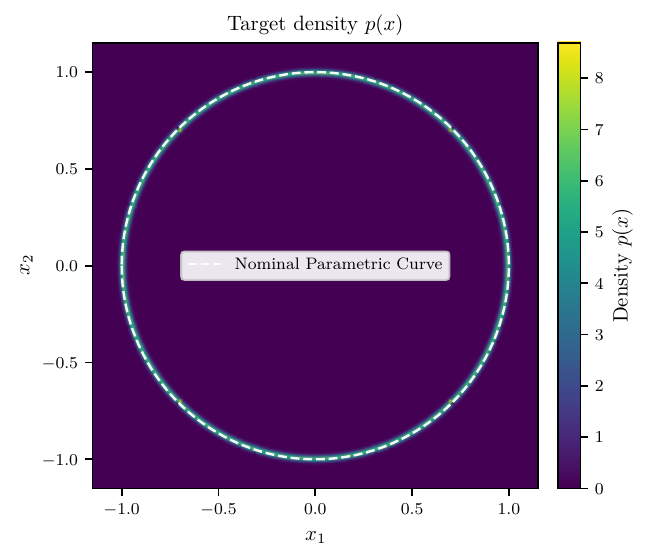}
        \caption{Target density $p(x)$.}
        \label{fig:target-density}
    \end{subfigure}
    \hfill
    \begin{subfigure}[t]{0.47\textwidth}
        \centering
        \includegraphics[width=\textwidth]
            {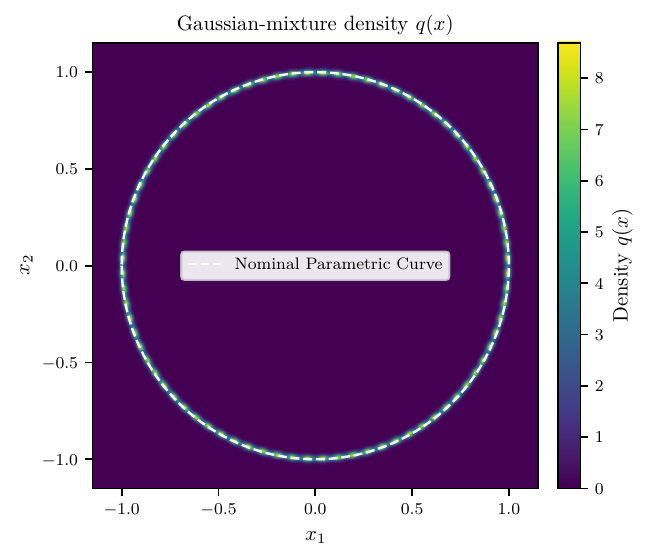}
        \caption{Gaussian-mixture density $q(x)$.}
        \label{fig:gaussian-density}
    \end{subfigure}

    \caption{Comparison of the target \ac{PDF} (uniform \ac{PDF}) and its
    \ac{GM} representation for the circular curve which is almost indistinguishable visually.}
    \label{fig:density-comparison}
\end{figure}
\begin{figure}[htbp]
    \centering
    \includegraphics[width=\textwidth]
        {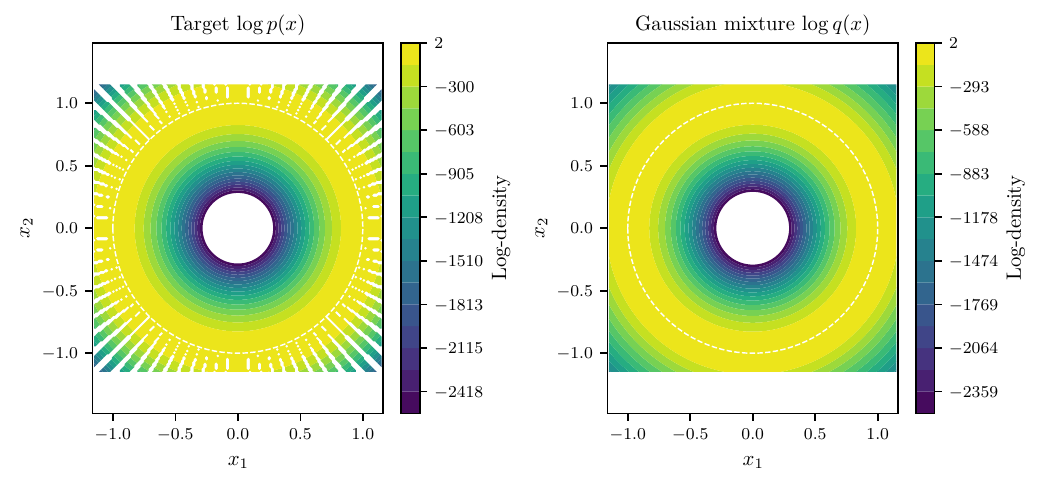}
    \caption{Comparison of the target log-density
    $\log p(x)$ (left) and the Gaussian-mixture
    log-density $\log q(x)$ (right) for the
    polygonal approximation of the unit circle.
    Both distributions are constructed using
    $M=80$ equally weighted segment components
    and normal noise standard deviation
    $\sigma_{\mathrm n}=0.01$.}
    \label{fig:circle-density-comparison}
\end{figure}
\FloatBarrier

Let $\gamma:[-\pi,\pi]\to\mathbb R^2$ denote the
unit-circle parametrization
\begin{equation}
    \gamma(t)
    =
    \begin{pmatrix}
        \cos t\\
        \sin t
    \end{pmatrix}.
    \label{eq:circle-parametrization}
\end{equation}
The corresponding continuous curve is
\[
    \Gamma
    =
    \left\{
        \gamma(t):t\in[-\pi,\pi]
    \right\}.
\]

To construct a simplicial approximation, we discretize
the parameter interval into $M$ equal subintervals.
Specifically, let
\begin{equation}
    t_j
    =
    -\pi+\frac{2\pi j}{M},
    \qquad j=0,\ldots,M,
    \label{eq:circle-parameter-grid}
\end{equation}
and define the corresponding vertices by
\begin{equation}
    v_j=\gamma(t_j),
    \qquad j=0,\ldots,M.
    \label{eq:circle-vertices}
\end{equation}
Since $\gamma(-\pi)=\gamma(\pi)$, we have
$v_0=v_M$. The resulting polygonal approximation
consists of the embedded $1$-simplices
\begin{equation}
    \sigma_j
    =
    \operatorname{conv}\{v_{j-1},v_j\},
    \qquad j=1,\ldots,M.
    \label{eq:circle-simplices}
\end{equation}

For each simplex, define its length, unit tangent,
and unit normal by
\begin{align}
    \ell_j
    &=
    \|v_j-v_{j-1}\|,\\
    t_j
    &=
    \frac{v_j-v_{j-1}}{\ell_j},\\
    n_j
    &=
    \begin{pmatrix}
        0 & -1\\
        1 & 0
    \end{pmatrix}t_j.
    \label{eq:circle-tangent-normal}
\end{align}
Because the vertices are equally spaced in the
angular parameter, all polygonal segments have
the same length,
\begin{equation}
    \ell_j
    =
    2\sin\left(\frac{\pi}{M}\right),
    \qquad j=1,\ldots,M.
    \label{eq:circle-segment-length}
\end{equation}

\Cref{fig:circle-approximation} illustrates the
continuous nominal curve and its polygonal approximation.

\begin{figure}[htbp]
    \centering
    \includegraphics[width=0.82\textwidth]
        {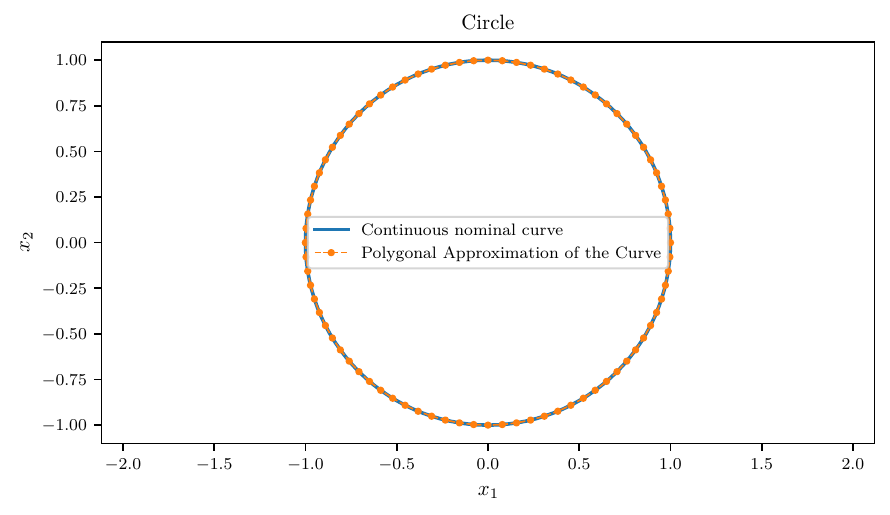}
    \caption{Unit-circle parametric curve and its polygonal
    approximation by \(M=80\) embedded \(1\)-simplices.
    The blue curve represents the continuous nominal
    parametric curve, while the orange polygon represents
    its piecewise-linear approximation.}
    \label{fig:circle-approximation}
\end{figure}
\FloatBarrier

We associate a probability distribution with each
embedded segment by introducing independent random
variables
\begin{equation}
    U_j\sim\mathcal{U}(u_j|0,1),\quad Z_j\sim\mathcal N(z_j|0,1),\quad U_j \perp Z_j.
\end{equation}
The corresponding \ac{RV} is defined as
\begin{equation}
    X_j
    =
    v_{j-1}
    +U_j(v_j-v_{j-1})
    +\sigma_{\mathrm n}Z_jn_j,
    \qquad j=1,\ldots,M,
    \label{eq:circle-component-model}
\end{equation}
where \(\sigma_{\mathrm n}>0\) denotes the standard
deviation of the Gaussian perturbation in the
normal direction.

Thus, the tangential coordinate is uniformly
distributed along the segment, whereas the normal
coordinate follows a centered Gaussian distribution.
The resulting component distribution is absolutely
continuous with respect to two-dimensional
Lebesgue measure.

Let \(p_j\) denote the density of \(X_j\).
For \(x\in\mathbb R^2\), introduce the local coordinates
\begin{align}
    s_j(x)
    &=
    t_j^\top(x-v_{j-1}),\\
    r_j(x)
    &=
    n_j^\top(x-v_{j-1}).
\end{align}
Then the component density is
\begin{equation}
    \label{eq:circle-component-density}
    p_j(x)=\frac{1_{[0,\ell_j]}(s_j(x))}{\ell_j\sqrt{2\pi}\sigma_{\mathrm n}}\exp\left(-\frac{r_j(x)^2}{2\sigma_{\mathrm n}^2}\right).
\end{equation}

We select the segment index according to the
arc-length weights
\begin{equation}
    w_j
    =
    \frac{\ell_j}{
        \sum_{k=1}^{M}\ell_k
    },
    \qquad j=1,\ldots,M.
    \label{eq:circle-arclength-weights}
\end{equation}
Since all segments have equal length in the present
construction, these weights simplify to
\begin{equation}
    w_j=\frac{1}{M}.
    \label{eq:circle-uniform-weights}
\end{equation}
The resulting target distribution is the mixture
\begin{equation}
    p(x)
    =
    \sum_{j=1}^{M}w_jp_j(x).
    \label{eq:circle-target-mixture}
\end{equation}

For each segment, the mean and covariance of \(X_j\)
are given by
\begin{equation}
    \label{eq:mean_var}
    m_j=\mathbb E(X_j)=\frac{v_{j-1}+v_j}{2},\quad\Sigma_j=\mathbb{V}(X_j)=\frac{\ell_j^2}{12}t_jt_j^\top+\sigma_{\mathrm n}^2n_jn_j^\top.
\end{equation}

By the componentwise information-projection result
established in this work, the unique forward-KL
optimal nondegenerate Gaussian approximation is
\begin{equation}
    q_j
    =
    \mathcal N(x|m_j,\Sigma_j).
    \label{eq:circle-component-gaussian}
\end{equation}
In particular, the Gaussian covariance captures
the uniform variation along the segment and the
independent Gaussian perturbation in its normal
direction.

For the uniform distribution on a \(1\)-simplex,
the minimum componentwise divergence is
\begin{equation}
    D_{\mathrm{KL}}(p_j:q_j)
    =
    C_1
    =
    \frac{1}{2}
    \log\left(\frac{\pi e}{6}\right)
    \approx 0.17649
    \quad\text{nats}.
    \label{eq:circle-component-kl}
\end{equation}
This quantity is independent of the segment length,
orientation, location, and normal noise variance.

We construct the Gaussian-mixture approximation
using the same component weights:
\begin{equation}
    q(x)
    =
    \sum_{j=1}^{M}w_jq_j(x).
    \label{eq:circle-gaussian-mixture}
\end{equation}
By joint convexity of the Kullback--Leibler
divergence, we obtain
\begin{equation}
    D_{\mathrm{KL}}(p:q)\leq\sum_{j=1}^{M}w_jD_{\mathrm{KL}}(p_j:q_j)\nonumber=\sum_{j=1}^{M}w_j C_1\nonumber=C_1=\frac{1}{2}\log\left(\frac{\pi e}{6}\right)\approx 0.17649\quad\text{nats}.
    \label{eq:circle-mixture-kl-bound}
\end{equation}
Consequently, the information projection divergence between the
target distribution and its \ac{GM}
divergence admits a bound that is independent
of the number of segments, the polygonal geometry,
and the normal noise variance.

The target log-density \(\log p(x)\) and the
Gaussian-mixture log-density \(\log q(x)\) are
evaluated and compared over the ambient
two-dimensional space.~\Cref{fig:circle-density-comparison}
shows the resulting log-density representations.
The left panel displays the target distribution,
whereas the right panel displays its
Gaussian-mixture approximation.

For the Monte Carlo simulation we estimate the \ac{KL} 
divergence \(D_{\mathrm{KL}}(p:q)\) using \(\#\text{Samples}=30000\). For
\(\#\text{Segments}=80\), \(\sigma_{\mathrm n}=0.01\), and random seed
\(42\), the estimated divergence is
\begin{equation}
    \widehat{D}_{\mathrm{KL}}(p:q)=0.045109,
\end{equation}
with standard error \(2.987\times10^{-4}\).
The corresponding \(95\%\) confidence interval is
\begin{equation}
    [0.044524,0.045695].
\end{equation}
This estimate lies strictly below the theoretical
upper bound
\begin{equation}
    D_{\mathrm{KL}}(p:q)\leq\frac{1}{2}\log\left(\frac{\pi e}{6}\right)\approx 0.17649.
\end{equation}
Thus, the numerical results are consistent with
the theoretical mixture-divergence guarantee and
indicate a substantially smaller empirical
divergence than the established upper bound.

The comparison illustrates how the individual
segment-wise Gaussian approximations combine to
produce a full-dimensional \ac{GM}
around the polygonal approximation of the
continuous circular curve. The resulting
distribution retains the overall geometric
structure of the target while replacing the
individual segment distributions by nondegenerate
Gaussian components.
\begin{remark}[General Curves]
\label{rem:generality-geometric-gaussian-mixtures}

Although the present example focuses on a circular
parametric curve, the underlying construction is by no
means restricted to this particular geometry. It applies
to general curves through their polygonal
approximations, with each segment inducing a
probabilistic geometric primitive and its corresponding
Gaussian component. In particular, the framework
accommodates smooth and closed curves, open curves,
non-regular curves, as well as self-intersecting
geometries and unions of curves.

This generality has already been demonstrated from a
computational perspective in the complementary work
\cite{darij_gm_curve,darij_gm_ieee}, 
where a broad collection of canonical plane curves 
is employed as test cases to investigate the 
expressive capabilities of geometric \ac{GM} representations. 
These experiments demonstrate that the resulting \ac{GM}
faithfully capture the local tangential and normal
structure, as well as the global geometry of the
underlying curves.

The present work complements this computational
investigation by providing the corresponding
mathematical foundations and theoretical guarantees
for the Gaussian approximation. In this sense, the
circular example considered here serves as a
representative instance of a substantially more
general geometric probabilistic construction, rather
than as a geometry-specific application.
\end{remark}
\begin{remark}
The example illustrates the application of the
componentwise Gaussian information projection
and the global \ac{GM} divergence bound to a
closed parametric curve. Although the underlying
continuous curve is nonlinear, its polygonal
approximation admits a direct representation
through embedded \(1\)-simplices.

Increasing the number of segments improves the
geometric resolution of the polygonal
approximation. Nevertheless, the theoretical
information projection upper bound remains unchanged because the componentwise divergence depends only on
the intrinsic dimension \(d=1\).

Furthermore, changing the normal \ac{PDF} standard
deviation modifies the spatial spread of the
target and \ac{GM} components without
affecting the theoretical bound.
\end{remark}
\begin{remark}[Connection to Embedded Simplicial Complexes]
\label{rem:circle-embedded-complex}

The polygonal approximation constructed above is a
one-dimensional geometric simplicial complex embedded
in \(\mathbb R^2\). In particular, its maximal simplices
are precisely the line segments
\(\{\sigma_j\}_{j=1}^M\).

This provides a concrete instance of the embedded
simplicial complexes introduced 
in~\Cref{def:embedded-simplicial-complex}.
Moreover, the segments intersect only at common
vertices, so the intersection property of~\Cref{prop:maximal-simplex-intersection}
is satisfied. Consequently, the polygonal realization
admits the decomposition into maximal simplices
required by the \ac{GM} theoretical upper bound
construction.

Thus, the example illustrates how a nonlinear
geometric object can be represented through a finite
embedded simplicial complex, allowing the theoretical
results to be applied directly.
\end{remark}
\begin{remark}[Uniform Distributions and Dirichlet Distributions]
\label{rem:circle-dirichlet}

The uniform distribution along each line segment
corresponds to the Dirichlet distribution with
parameter vector \(\alpha=(1,1).\) Indeed, the barycentric coordinates of a point on
a \(1\)-simplex are distributed according to
\(\mathcal{D}(x|1,1)\), which is simply the
uniform distribution on the standard \(1\)-simplex.

The component distributions in
\eqref{eq:circle-component-model} are therefore
special cases of the Dirichlet--Gaussian
construction introduced in~\Cref{def:dirichlet-gaussian}.

Furthermore, the component means and covariances
in~\Cref{eq:mean_var} are precisely
the uniform-simplex moment formulas established
in~\Cref{prop:embedded-dirichlet-moments}.

Hence, the numerical construction is not specific
to the circular geometry but is an instance of the
general Dirichlet--Gaussian framework developed
in the preceding sections.
\end{remark}
\begin{remark}[Componentwise Gaussian Information Projection]
\label{rem:circle-information-projection}

The Gaussian components in
\eqref{eq:circle-component-gaussian} are not chosen
heuristically. They are the unique information projection of the corresponding distributions onto the Gaussian family, as established by~\Cref{thm:component-projection}.

In particular, the Gaussian means and covariance
matrices are determined entirely by moment matching.
The covariance decomposition
\begin{equation}
  \Sigma_j=\underbrace{\frac{\ell_j^2}{12}t_jt_j^\top}_{\text{tangential variation}}+\underbrace{\sigma_{n}^2n_jn_j^\top}_{\text{normal perturbation}}
\end{equation}
makes the geometric interpretation of the
information projection explicit.

The first term captures the uniform distribution
along the segment, whereas the second preserves
the Gaussian \ac{PDF} introduced in the normal
direction. This example therefore illustrates
how the abstract information projection result
translates into a geometrically interpretable
Gaussian approximation.
\end{remark}
\begin{remark}[Normal Gaussian Independence]
\label{rem:circle-noise-independence}

An important consequence of the componentwise
information projection formula is that the optimal 
approximation error does not depend on the normal \ac{PDF} 
variance \(\sigma_{\mathrm n}^2\).

Although changing \(\sigma_{\mathrm n}\) modifies
the spatial spread of both the target components
and their information projection onto the Gaussian family, 
the normal Gaussian contribution cancels exactly in the
componentwise information projection. Consequently,
\begin{equation}
    D_{\mathrm{KL}}(p_j:q_j)=\frac12\log\left(\frac{\pi e}{6}\right)\approx 0.17649
\end{equation}
for every segment and every \(\sigma_{\mathrm n}>0\).

This illustrates the general independence of
the normal covariance operators established 
in~\Cref{rem:covariance_operator_tanget_normal}
and shows that the \ac{KL} divergence is
governed by the intrinsic Dirichlet--Gaussian distribution 
on the embedded simplex rather than the ambient 
Gaussian \ac{PDF}.
\end{remark}
\begin{remark}[Global Mixture Guarantee]
\label{rem:circle-global-mixture-guarantee}

The construction of the full \ac{GM}
in~\Cref{eq:circle-gaussian-mixture} illustrates
the passage from componentwise information
projections to a \ac{GM} with a sharp upper bound.

Since the \ac{GM} weights are identical for the
target and approximating distributions, joint
convexity of the \ac{KL} divergence,
as established in~\Cref{prop:joint-convexity}, gives
the bound in~\Cref{thm:dirichlet-mixture-bound}.
For the present example, the intrinsic dimension
is \(d=1\), and the resulting upper bound is
\begin{equation}
    D_{\mathrm{KL}}(p:q)\leq C_1=\frac12\log\left(\frac{\pi e}{6}\right)\approx 0.17649.
\end{equation}
Importantly, this guarantee is independent of
the number of segments \(M\), their lengths, their
orientations, and the normal \ac{PDF} variance.
Thus, increasing the number of segments improves
the geometric resolution of the polygonal
approximation without changing the theoretical
sharp upper bound on the \ac{GM} divergence. 
As emphasized in~\Cref{rem:mixture-bound-not-global-optimality},
the result establishes a sharp upper bound
for the constructed \ac{GM}, rather
than global optimality over all possible
\acp{GM}.
\end{remark}
\end{example}
\section{Conclusion}
\label{sec:conclusion}

We derived the unique Gaussian information
projection of Dirichlet--Gaussian \acp{PDF} on embedded simplices,
obtaining a closed-form minimum information projection which is invariant up to 
an affine isomorphism and independent of the ambient dimension and
normal covariance. By joint convexity of the \ac{KL} divergence, 
we further established the sharp intrinsic-dimension-dependent \ac{KL} divergence 
upper bounds for finite \acp{GM}, giving sharp approximation guarantees
for \acp{GM} associated with embedded simplicial
complexes. A numerical example on a circular curve illustrates the 
theoretical results and their applicability to embedded manifolds represented by simplicial complexes.
\section{Acknowledgment}
This work was funded by Project SFB~1574/CRC~1574 ``A Circular Factory for the Perpetual Product'', 
Deutsche Forschungsgemeinschaft (DFG), Project ID 471687386.
\printbibliography

@book{info_geo,
  title     = "{Information Geometry}",
  author    = "{Nihat Ay} and {Jürgen Jost} and {Hông Vân Lê} and {Lorenz Schwachhöfer}",
  publisher = "Springer",
  year      =  2017,
}

@book{stuart_inverse,
  title     = "{Inverse Problems and Data Assimilation}",
  author    = "{Daniel Sanz-Alonso} and {Andrew Stuart} and {Armeen Taeb}",
  publisher = "Cambridge University Press",
  year      =  "2023",
}

@book{munkres_alg_top,
  title     = "{Elements of Algebraic Topology}",
  author    = "{James R. Munkres} and {Steven G. Krantz} and {Harold R. Parks}",
  publisher = "{CRC Press}",
  edition   =  2,
  year      =  2025,
}

@book{lee_topol,
  title     = "{Introduction to Topological Manifolds}",
  author    = "{John Marshall Lee}",
  publisher = "Springer",
  edition   =  2,
  year      =  2011,
}

@book{evans_measure,
  title = {Measure Theory and Fine Properties of Functions},
  edition = {2},
  publisher = {CRC Press},
  author = {{Lawrence C. Evans} and {Ronald F. Gariepy}},
  year = {2025}
}

@book{bernstein_matrix,
  title     = "{Scalar, Vector, and Matrix Mathematics}",
  author    = "{Dennis Stanley Bernstein}",
  publisher = "Princeton",
  year      =  2018,
}

@misc{darij_simplex,
  title         = "Closed-form Gaussian Mixture representations of
                   d-dimensional triangulated domains",
  author        = "Darijani, Ali and Stratmann, Benedikt and Hanebeck, Uwe D
                   and Thapa, Ratan Bahadur and Beyerer, J{\"u}rgen and Afifi,
                   Nehal",
  year          =  2026,
  copyright     = "http://arxiv.org/licenses/nonexclusive-distrib/1.0/",
  archivePrefix = "arXiv",
  primaryClass  = "cs.IT",
  eprint        = "2611.00922",
  note = {Accepted at 2026 IEEE Sensor Data Fusion: Trends, Solutions, Applications (SDF 2026)},
}

@book{dirichlet_tang,
title     = "{Dirichlet and Related Distributions}",
  author    = "{Kai Wang Ng} and {Guo-Liang Tian} and {Man-Lai Tang}",
  publisher = "Wiley",
  year      =  "2011",
}

@article{darij_gm_ieee,
  title     = "A Geometric Gaussian Mixture Representation of Curves and
               Trajectories in Euclidean Space",
  author    = "Stratmann, Benedikt and Soufian, Emadedin and Beyerer,
               J{\"u}rgen and Darijani, Ali",
  publisher = "Institute of Electrical and Electronics Engineers (IEEE)",
  note         = {Accepted at 2026 IEEE International Conference on Multisensor Fusion and Integration for Intelligent Systems (MFI 2026)},
  year      =  2026
}

@article{darij_gm_curve,
  author        = {Darijani, Ali and Stratmann, Benedikt and Beyerer, Jürgen},
  title         = {A Geometric Gaussian Mixture Representation of Plane Curves},
  journaltitle  = {arXiv preprint arXiv:2606.06505},
  year          = {2026},
  eprint        = {2606.06505},
  eprinttype    = {arxiv},
  eprintclass   = {cs.CG},
  doi           = {10.48550/arXiv.2606.06505},
  note         = {Accepted at 2026 IEEE International Conference on Multisensor Fusion and Integration for Intelligent Systems (MFI 2026)}
}

@book{nonparam_alex,
  title = {Introduction to Nonparametric Estimation },
  editor = {},
  publisher = {Springer},
  author = {Alexandre B. Tsybakov},
  year = {2009}
}
\end{document}